\documentstyle[12pt]{article}
\title{Integrability in $3+1$ Dimensions:\\Relaxing a Tetrahedron Relation}
\author{I.G.~Korepanov \thanks{South Ural State University, 76 Lenin av.,
Chelyabinsk 454080, Russia (permanent address), and Ural Division of
International Institute for Nonlinear Research, Ufa, Russia.}}
\date{December 1997}
\def\be{\begin{equation}}
\def\ee{\end{equation}}
\def\Dc{{\cal D}}
\begin{document}
\maketitle

\begin{abstract}
I propose a scheme of constructing classical integrable models in
$3+1$ discrete dimensions, based on a relaxed version of the problem of
factorizing a matrix into the product of four matrices of a special form.
\end{abstract}

It is well known that in the $2+1$-dimensional case the integrability
of classical models is closely connected with the problem of
re-factorizing the product of three special block matrices in a similar
product, but taken in the reverse order~\cite{korepanov dis,sergeev
ferroelectro,FTE}. It seems that all the known classical
integrable systems can be obtained from
there as some particular and/or limit cases~\cite{kashaev electro,sergeev
ferroelectro}. Moreover, the quantization of at least some systems
arising from this re-factorizing problem is amazingly
straightforward~\cite{FTE}.

The similar problem for the product of {\em four\/} matrices
(``tetrahedron relation'', see below) cannot, generally, be solved,
because of lack of parameters. However, it turns out that this problem
can be relaxed in such a way that it becomes solvable, and many
solutions of the ``evolution equations'' of the corresponding dynamical
system can be written out in algebraic-geometrical form, namely
employing an algebraic curve and divisors in it.

By the ``tetrahedron relation'' I mean in this paper the following
decomposition of a $6\times 6$ matrix~$K$ of complex numbers:
\be
K=A_{123}B_{145}C_{246}D_{356},
\label{relaxing *}
\ee
where the subscripts show in which lines and columns a matrix can have
nontrivial elements, while all other elements are zeroes if they are
non-diagonal and unities if they are diagonal. For example,
$$
A_{123}=\pmatrix{
a_{11} & a_{12} & a_{13} & 0 & 0 & 0 \cr
a_{21} & a_{22} & a_{23} & 0 & 0 & 0 \cr
a_{31} & a_{32} & a_{33} & 0 & 0 & 0 \cr
  0    &   0    &    0   & 1 & 0 & 0 \cr
  0    &   0    &    0   & 0 & 1 & 0 \cr
  0    &   0    &    0   & 0 & 0 & 1
}.
$$
We will also write $A$ instead of $A_{123}$ and so on.

The matrix $K$ has 36 parameters, while the product $ABCD$ has only
30 of them (4 matrices each having 9 parameters, but 6 of those
must be subtracted because of the obvious ``gauge'' freedom
in $A$, $B$, $C$ and $D$ not changing $K$, compare Chapter~2
of~\cite{korepanov dis}). So let us relax the relation~(\ref{relaxing *})
in the following way. Let $\Phi_k$ and $\Psi_k$, where $k=1,2,3,4,5$,
be five given six-dimensional vectors, and let us search for such
matrices $A$, $B$, $C$ and $D$ that
\be
ABCD\,\Phi_k=\Psi_k
\label{relaxing **}
\ee
for all $k$. The conditions (\ref{relaxing **}) impose exactly 30
restrictions on $A$, $B$, $C$ and $D$, as needed.

We can depict each of the matrices $A$, $B$, $C$ and $D$ as a vertex
with three ``incoming'' and three ``outgoing'' edges. Then, the product
$ABCD$ will be represented as a tetrahedron with six ``inner'' edges
and twelve ``outer'' edges. Six of the latter are incoming, and to them
the components of vectors $\Phi_k$ are attached, and six others are
outgoing, and to them the components of vectors $\Psi_k$ are attached.
Now, the fact that we can find $A$, $B$, $C$ and $D$ from
(\ref{relaxing **}) means that the numbers attached to the {\em inner\/}
edges are also determined from $\Phi_k$ and $\Psi_k$ (up to the gauge
freedom: we can attach to each edge an arbitrary constant and multiply
all the edge's five numbers by that constant).

This suggests to introduce the following dynamical system, in the spirit
of~\cite{FTE}. Let there be several planes in the three-dimensional space.
Let them be permitted to move arbitrarily, but each remaining parallel
to its initial position. We will call {\em edges\/} of such figure
the (open) intervals of lines where two planes intersect (we will attach some
orientation to each of those lines) between points
({\em vertices\/}) where three planes intersect. To each edge five complex
numbers (``dynamical variables'') will correspond, and all those numbers
must be consistent
in the sense that there must exist $3\times 3$ matrices in each vertex
transforming five incoming 3-vectors into five outgoing ones.
As already explained, for just one tetrahedron this consistency
means that the variables on inner edges are expressed through those
on outer edges, and the reader can invent more examples where also
the variables can be chosen freely for some edges and then all the other
variables get fixed.

The elementary move in the system can be described as passing a plane through
the intersection point of three other planes, or ``turning a tetrahedron
inside out''. In such move, the numbers corresponding to the tetrahedron's
outer edges remain intact, while those corresponding to the inner edges
change according to the change of ``relaxed
decomposition''~(\ref{relaxing **}), e.g., instead of~(\ref{relaxing **})
we get
$$
D'C'B'A'\,\Phi_k=\Psi_k.
$$

An algebro-geometric solution for this system reads as follows.
Let us take an algebraic curve~$\Gamma$, and attach to each plane,
say plane number~$j$, two divisors $\Dc_j^{(1)}$ and $\Dc_j^{(2)}$
in~$\Gamma$ of degree~2. For simplicity,
we can assume that such a divisor is just two arbitrary points of~$\Gamma$.
To be exact, we will attach $\Dc_j^{(1)}$ to one open half-space in which
plane number~$j$ divides the space, and $\Dc_j^{(2)}$---to the other
open half-space, and nothing to the plane itself.

Let us also take some divisor $\Dc$ whose degree will be specified later,
and put in correspondence to each point~$x$ of the space the divisor
$$
\Dc(x)=\Dc-\sum_j \Dc_j^{(\alpha_j)},
$$
where the sum is taken over all such $j$ that $x$ does not belong
to the plane number~$j$, and $\alpha_j=1$ or $2$ according to which
half-space the point~$x$ belongs.

We will be interested, as in~\cite{FTE}, in divisors corresponding this way
to the {\em edges}. Let us choose the degree of $\Dc$ so that the degree
of $\Dc(x)$, if $x$ belongs to an edge, be~$g$---the genus of~$\Gamma$.
If each $\Dc_j^{(\alpha_j)}$ is just two points, this means that the
space of meromorphic functions with poles in the points of~$\Dc$
and zeroes in points of all $\Dc_j^{(\alpha_j)}$ is one-dimensional,
that is, to each edge corresponds a meromorphic function~$f$,
determined up to a constant multiplier.
Now let us choose five more points $z_1,z_2,z_3,z_4,z_5\in\Gamma$
(the same for all edges), and put in correspondence to the edge
the numbers $f(z_1),f(z_2),f(z_3),f(z_4)$ and $f(z_5)$.
The same reasoning as in \cite{korepanov dis,FTE} shows that this
provides a solution for our dynamical system.

\end{document}